# Born In Bradford Mobile Application

Stella Lee
Department of Computing
University of Bradford
Bradford BDF 1DP, UK

S.E.A.Lee@bradford.ac.uk,
stellaleeuss@gmail.com

Martin Walda
Department of Computing
University of Bradford
Bradford BDF 1DP, UK

M.J.Walda@bradford.ac.uk

Vasiliki Delimpasi
Department of Design
University of Bradford
Bradford BDF 1DP, UK

V.Delimpasi@student.bradford.ac.uk

## ABSTRACT
The Born In Bradford mobile application is an Android mobile application and a working prototype that enables interaction with a sample cohort of the Born in Bradford study. It provides an interface and visualization for several surveys participated in by mothers and their children. This data is stored in the Born In Bradford database. A subset of this data is provided for mothers and children.

The mobile application provides a way to engage the mothers and promote their consistency in participating in subsequent surveys. It has been designed to allow selected mothers to participate in the visualization of their babies' data. Samsung mobile phones have been provided with the application loaded on the phone to limit and control its use and access to data.

Mothers login to interact with the data. This includes the ability to compare children's data through infographics and graphs and comparing their children's data with the average child. This comparison is done at different stages of the children's ages as provided in the dataset.

## Categories and Subject Descriptors
D.3.3 [**Programming Languages**]: Java Android Mobile Development – Data Constructs and Features – *threading, control structures, SOAP Web Services.*

## 1. INTRODUCTION
Born in Bradford carries out a lot of surveys by working with families and carrying out surveys on children born in those families. This is aided by the help of community groups and health professionals to help detect common traits that lead to children having an unhealthy life. The data collected about babies born in Bradford is periodic and the child growth data collected from surveys has created a platform of observation for health researchers. The information populated in Born In Bradford's database is used to improve the health of children in Bradford. This has provided a platform to discern why some children are healthy and why others remain sick.

Mothers involved in this study are presented a community of people to help guide the health of their children and take appropriate measures to improve the health of their children. Mothers are exposed to advice that can improve the diet of their children and family as a whole. Mothers develop better insight in the process of being exposed to the healthy child development programme and become aware of the genetic factors and other factors that could limit the health of their children. These mothers have given the Born In Bradford project consent to access their medical records and their children's records. These records include bio samples, height and weight measurements.

## 2. Visualization of Child Data
Born In Bradford (BiB) carries out surveys on children from time to time. This has stirred up questions about the use of the data and why mothers cannot view this data or manage the profiles of their children. Mothers desire to know the health status of their children and be kept up to date with information on their children from the multiple surveys they have participated in. The heights and weights of children are recorded often. These mothers would like to see a comparison of their children and the average child to verify that their children are growing properly. The easiest visualization a mother can have is a mobile application that provides appealing visualization to mothers. This would need to be accessible on mothers' phones. The Born In Bradford team have proposed a prototype of the application for the mothers to interact with their children's data hence they have chosen some mothers and children for this exercise and have provided a subset of the Born In Bradford data which has been restricted to the chosen mothers and children. These mothers have been provided with Android mobile phones for this exercise to restrict the application development, test and use of the mobile application to Born In Bradford's environment and not Google Play. Hence mothers would not be able to download the application but have it preinstalled on the phones presented to them.

### 2.1 Architecture
The architecture of the BiB mobile application comprises of a server back end which hosts a MySQL database and Java Simple Object Access Protocol (SOAP) web services that interact with the database on behalf of the mobile application user. SOAP is an XML-based protocol that can be carried over any transport mechanism capable of delivering a byte stream. In practice, SOAP messages are usually exchanged between clients and services that are resident in web containers and are typically encapsulated inside an HTTP request or response message [1]. The web service implements a layer of abstraction in the design architecture of the BiB mobile application solution. The level of abstraction consists of the retrieval of data from the database and implementation of some business logic to reduce the level of coupling in the design of the solution. This will provide a modular approach to debugging the solution and maintenance of the solution as a whole.

The current architecture designed for this project is a closed architecture which will not be hosted on Android Google Play but will be maintained as a prototype for future development of the BiB mobile application service. The Born In Bradford organization which is the primary support for the BiB data, hosts a large database of data for the mothers and children involved in the Born In Bradford research and due to security protocols, provides a subset of the data in an anonymous format. This means that the association between mothers and children is via unique IDs and not formal names which can be recognized. This data is manually provided to the University of Bradford and hosted at the University's premise. The design of this solution is a no update system hence, the BiB main database server is not updated. This data is sent either through a flash drive or an email after sign off between the BiB team and the University and is updated subsequently through this means.

Web services interact with the BiB dataset hosted at the University of Bradford's premise and exposes a single end point to the BiB mobile application via KSOAP interface. KSOAP is an interface that enables interaction between SOAP web services and Android devices. KSOAP is a SOAP API based on kXML, where kXML is a lightweight Java-based XML parser designed to run on limited, embedded systems such as personal mobile devices [2] KSOAP does not support the entire SOAP specification. It supports most of the commonly used SOAP features and is sufficient for most web services that are currently available. Currently, almost all major KSOAP applications and tools are based on KSOAP release v1.2, which supports a core subset of SOAP features [3].

A mobile application user which is a mother, requests for a service and the application in turn requests for data or updates data on the phone via a web service call. The web service requests data from the database and updates the android device via the KSOAP interface library. The application attempts to load data from the web service when the application is installed and used for the first time and stores the data on the mobile phone. On subsequent use of the application, the application checks for updates to the dataset via a web service call by comparing the last update date of the data on the phone and the current update date in the database. Once a change is noticed after comparing the dates, the mobile application deletes the previous data on the mobile phone and performs an update on the mobile phone. This update occurs as frequently as update to the data is performed.

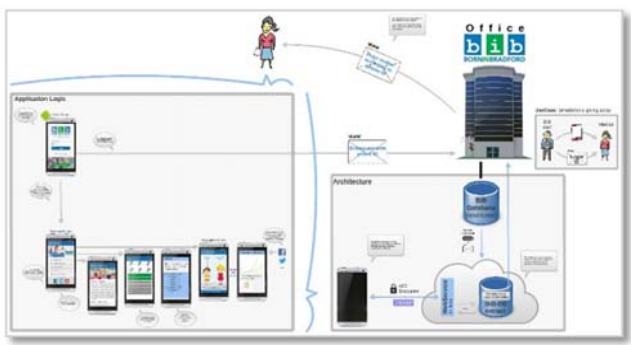

**Figure 1. BiB Mobile Application Architecture**

The architecture also takes into account forgotten passwords which is requested via a link on the login user interface. This request is sent to the BiB team who respond to the mother's request by physically handing the mother her ID. The BiB team adopted this approach because they were uncertain about the security of the mother's data. This functionality would be reviewed in subsequent development versions which would involve changes to the current architectural design of the mobile application.

## User Interface Design

The login view has been proposed to signify the authentication of a mother on the BiB mobile application. The view has an input for entering a unique token id which will be used by a mother for authentication. The login button is very visible to make the login process simple to understand and use.

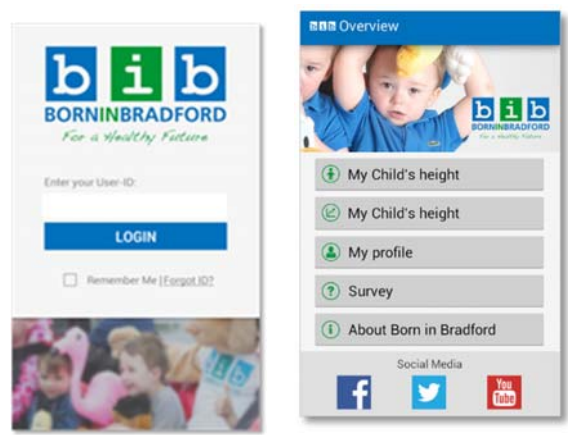

**Figure 2. Login and Overview View**

The login view displays some other functionality which are the Remember Me and the Forgot ID functionality. These authentication enablers are standard web login functionalities which can easily be understood.

The overview view was proposed to present mothers with a main menu functionality which displays the different options available for a mother to choose. The design is very simple with links at the bottom to social medias. The overview page will be displayed after the mother has been successfully authenticated.

The pictorial child view will display a child's height which can be compared against the national average child. The design consists of two silhouettes: a green colored silhouette which represents a mother's child and a grey colored silhouette which represents the national average. There are two labels at the top of the silhouettes which represents the height of the children in centimeters. A progress bar exists at the bottom of the silhouettes. When a mother slides the progress bar, the height of the child increases in response to the position of the highlighted part of the progress bar. There is an option to select different children and compare their heights. This applies to mothers with more than one child. The design also has a share symbol at the top right side of the page to enable a mother share a screen shot of her child's height at a point on the progress bar to any social media installed on the phone.

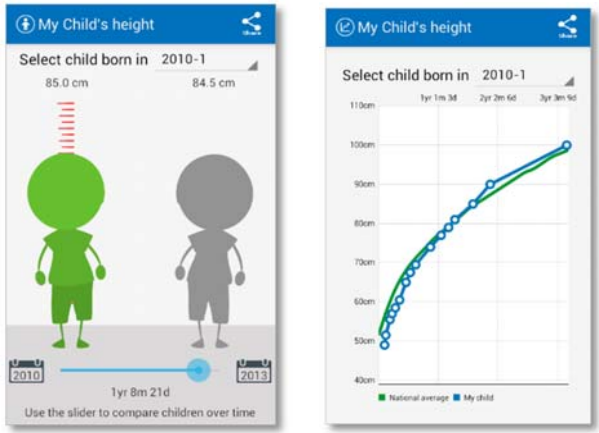

**Figure 3. Pictorial and Graphical Child Height View**

The graphical child view is similar to the pictorial child view. Unlike the pictorial child view, it will display a child's height which can be compared against the national average child in a graphical format. The graph is plotted as height against time. The blue line represents a mother's child and the green line represents the national average. The design has the share option to enable a mother share her child's information and if she has more than one child, she has the option of selecting different children and she can view their heights over time.

## 3. Prototype Enhancements
## 3.1 Mobile Diagnosis

Gittens [4] explains that recent studies have suggested that the implementation of various mobile health technologies help combat diabetes and other non-communicable diseases. According to the World Health Organization, there is evidence to support that diabetes cases and their related complications can be averted by having a healthy diet, engaging in regular physical activity and maintaining normal body weight.

The mobile application could provide a platform for intelligent diagnosis and advice systems that can help mothers plan their children's diet based on their current weight, height and other information gotten from the mothers via mobile surveys. Mothers can have smart reminders for children's meals, water and BiB meetings. Change in diet may be advised for a child if the family of the child have the tendency to inherit some health conditions such as a heart condition or diabetes.

The application can be extended to provide mobile diagnosis services. The mobile application can synchronize with children's full medical record instead of a subset of their medical record. The mobile application would represent a child before a General Practitioner (GP). This will be possible after synchronizing the health system and mobile application. The GP could be sent urgent medical requests and medical advice could be given to parents to handle minimal health condition such as running temperature at night, booking an appointment to see the doctor or an emergency. In the case of an emergency, the system would know the child or home involved due to its access to the mothers address and can be extended to get GPS locations which enables medical personnel locate the home of the child.

Ucan and Gu [5] proposed a platform for developing privacy preserving diagnosis mobile applications by enforcing a framework that consists of three parts: Natural Language Processing – which extracts symptoms from the user's free text input; Privacy Preserving Information Retrieval – privacy preserving database lookup; and Decision Support-rule based decision support. The system supports mobile application by taking symptoms from users, sending back possible diagnosis to the users, and hiding users' medical information.

The mobile application will contain a questionnaire with basic questions for symptoms and a free text field for a mother to express herself generally about the current condition of the child. These symptoms give the GP basic knowledge of the child's health. In the case of the child coming over to the hospital, the child's symptoms are verified and prescriptions are made for the child.

For this system to be seamless and automated the NHS would need to review its system to expose services that could be plugged into the mobile application to provide an integrated system and this would require that the Born In Bradford team, a branch of the NHS provide the mothers' records to the NHS health system and the anonymity of the mothers would need to be removed by using a map to assign current mothers their respective identifiers on the NHS system.

## 4 Child and Mother Profile Customization

Mothers made a request to update their children's profile pictures or customize their silhouette in one of the user acceptance test they were involved in. They requested for the customization to be made based on their nationality and ethnicity. Mothers who were Asian wanted a silhouette that looked Asian with various customizations to look like their children. They requested that they be given the ability to do the customizations. This is more of implementing something fun that mothers can do on the platform apart from participating in the BiB research.

Ni, Jin And Jiang [6] explains the importance of graphics engine in managing graphics. Graphics engine is the main program that is used to control all games' graphics function and also is the game engine. It bundles all the graphics elements together and manage their work orderly in the background. A good graphics engine will not only reduce the duplication of codes, but also lessen the complexity of applications development. And most important is that it can improve the developing efficiency.

In order to implement profile graphics customization functionality, it is advisable to implement a graphics engine that manages the customization process. This is because a child silhouette could be customized on different levels from color change to customization of clothes, hair, shoes and many more. There are many possible variations of combination of graphics and the ability of the mobile application to use graphics without duplicating them or duplicating code would increase the efficiency and speed of the application. The graphic engine needs to be developed in a way that it handles system resources efficiently and not use up all the memory on the mobile device.

## 5 Conclusion

The Born In Bradford Android mobile application implementation of the BiB research, is a prototype that can provide more services to the mothers involved in the research and also reduce the cost involved in the research by providing other means and platforms of interactions with the mothers and their children. This paper described the architecture of the mobile application and the possibility of introducing consultation, more flexibility and a workflow that allows mothers interact and update the BiB Dataset and subsequently the Born In Bradford universal data.

## 6 ACKNOWLEDGEMENTS

Special thanks to Professor Mick Ridley, who supervised this writing, the Born In Bradford team for their availability to guide in the development of the mobile application and providing expertise and data that makes the mobile application useful, Simon Couth, Andrew Holmes, and David Robison for excellent project management of the Born In Bradford Mobile Application prototype development.